\documentclass[11pt]{article}

\usepackage{amsmath}
\usepackage{amssymb}
\usepackage{amsfonts}

\def\Mendonca{Mendon\c{c}a}
\def\C{\mathcal{C}}
\def\D{\mathcal{D}}
\def\d{\mathrm{d}}

\def\H{\mathcal{H}}
\def\R{\mathcal{R}}
\def\ave#1{\left\langle#1\right\rangle}
\def\abs#1{\left|#1\right|}

\begin{document}
\begin{center}

\textbf{\LARGE Quantum gravitational decoherence of matter waves}

\vspace{3mm}

Charles H-T Wang$^{1,2,a}$, Robert Bingham$^{2,3,b}$ and J Tito \Mendonca{}$^{4,2,c}$

\vspace{3mm}

\begin{small}
$^1$School of Engineering and Physical Sciences,\\
University of Aberdeen, King's College, Aberdeen AB24 3UE, UK

\vspace{1mm}

$^2$Rutherford Appleton Laboratory, Chilton, Didcot, Oxon OX11 0QX, UK

\vspace{1mm}

$^3$(SUPA) Department of Physics,\\ University of Strathclyde, Glasgow G4 0NG, UK

\vspace{1mm}

$^4$CFP, Instituto Superior Tecnico, 1049-001 Lisboa, Portugal

%\email{c.wang@abdn.ac.uk}
%\affiliation{SEPS, University of Aberdeen, King's College, Aberdeen AB24 3UE, UK}
%\affiliation{Rutherford Appleton Laboratory, Chilton, Didcot, Oxon OX11 0QX, UK}

%\email{r.bingham@rl.ac.uk}
%\affiliation{Rutherford Appleton Laboratory, Chilton, Didcot, Oxon OX11 0QX, UK}
%\affiliation{(SUPA) Department of Physics, University of Strathclyde, Glasgow G4 0NG, UK}

%\email{titomend@ist.utl.pt}

\vspace{2mm}

$^a$c.wang@abdn.ac.uk, $^b$r.bingham@rl.ac.uk, $^c$titomend@ist.utl.pt

\vspace{2mm}


\end{small}
\end{center}

%\noindent

\begin{abstract}
One of the biggest unsolved problems in physics is the unification
of quantum mechanics and general relativity. The lack of
experimental guidance has made the issue extremely evasive, though
various attempts have been made to relate the loss of matter wave
coherence to quantum spacetime fluctuations. We present a new
approach to the gravitational decoherence near the Planck scale,
made possible by recently discovered conformal structure of
canonical gravity. This leads to a gravitational analogue of the
Brownian motion whose correlation length is given by the Planck
length up to a scaling factor. With input from recent matter wave
experiments, we show that the minimum value of this factor to be
well within the expected range for quantum gravity theories. This
suggests that the sensitivities of advanced matter wave
interferometers may be approaching the fundamental level due to
quantum spacetime fluctuations and that investigating Planck scale
physics using matter wave interferometry may become a reality in
the near future.
\end{abstract}

%\section{Introduction}

Physics on the large scale is based on Einstein's theory of
general relativity (GR), which interprets gravity as the curvature
of spacetime. Despite its tremendous success as an isolated theory
of gravity, GR has proved problematic in integration with physics
as a whole, in particular the physics of the very small governed
by quantum mechanics. There can be no unification of physics,
which does not include them both.  Superstring
theory~\cite{Green1987} and its recent extension to the more
general theory of branes is a popular candidate, but the links
with experiment are very tenuous. Loop quantum
gravity~\cite{Ashtekar1986,RovelliSmolin1995c} attempts to
quantize GR without unification, and has so far received no
obvious experimental verification.

One hundred years ago, when Planck introduced the constant named
after him, he also introduced the Planck scales, which combined
this constant with the velocity of light $c$  and Newton's
gravitational constant $G$ to give the fundamental Planck time
$T_\text{Planck}=(\hbar\,G /c^5)^{1/2} \approx 10^{-43}$~s,
Planck length $L_\text{Planck}=c\,T_\text{Planck}\approx10^{-35}$~m
and Planck mass $M_\text{Planck}={\hbar}/({c^2T_\text{Planck}})\approx10^{-8}$~kg.
Experiments on quantum
gravity require access to these scales.  To access these scales
directly using accelerators would require $10^{19}$~GeV
accelerators, well beyond any conceivable experiments.

Nonetheless, there have been significant developments in examining
possible signatures of quantum gravity. A number of phenomena have
been proposed to be potentially observable, including
`deformations' of special relativity and `imprints' in the cosmic
microwave background. For recent reviews, see e.g.
\cite{Amelino-Camelia2005a} and \cite{Smolin2006} and references
therein.

One possible way of accessing the Planck scale is to use the
concept developed by Einstein in his study of thermal fluctuations
of small particles through the Brownian
motion~\cite{Einstein1905}.  Modern experimental methods are so
much in advance of those of Einstein's time, that we are now in a
position to consider accessing the Planck scales by a method
analogous to Brownian motion, a concept first pointed out by Ellis
et al~\cite{Ellis1984}. The curvature of spacetime produces
changes in proper time, the time measured by moving clocks. For
sufficiently short time intervals, near the Planck time, the
proper time fluctuates strongly due to quantum fluctuations.  For
longer time intervals, proper time is dominated by a steady drift
due to smooth spacetime. Proper time is therefore made up of the
quantum fluctuations plus the steady drift.  The boundary
separating the shorter time scale fluctuations from the longer
time scale drifts, is marked by a time
$\tau_0=\lambda\,T_\text{Planck}$ that defines the borderline
between semiclassical and fully quantum regimes of gravity, in
terms of a dimensionless cut-off parameter $\lambda$.

%%% new
Since the quantum to classical transition of gravity is expected
to occur at a length scale $\lambda_0=\lambda\,L_\text{Planck}$
larger than $L_\text{Planck}$ by a few orders
of magnitude, the parameter $\lambda$ should approximately satisfy
$\lambda > 10^2$. Its actual value is model dependent. E.g. the
extra-dimensional gravity model as per Arkani-Hamed et al has the
fundamental scale of gravity in the TeV range
\cite{Arkani-Hamed1998}, which would place $\lambda$ to be as
large as around $10^{16}$. Below, the motion of a quantum wave
packet will be examined on a length scale larger than $\lambda_0$
where the time-varying background metric is assumed to be well
described as a classical field. Markopoulou et al have recently
developed a theoretical framework where particles are described as
`noiseless subsystems' in a background independent manner
\cite{Markopoulou2005}. However, we assume that at an appropriate
low-energy limit of such a theory, the classical picture of
geometry will emerge.
%%%

Matter wave interferometers are ideal in measuring decoherence
effects and will be able to put upper limits on quantum
fluctuations which will help guide the theoretical work. An atom
is a quantum clock with a very high frequency proportional to its
mass. In contrast, the proper time intervals of massless particles
like photons are unaffected as they travel along null geodesics.
In an atom interferometer, an atomic wavepacket is split into two
coherent wavepackets that follow different paths before
recombining.  The phase change of each wavepacket is proportional
to the proper time along its path, resulting in constructive or
destructive interference when the wavepackets recombine.
The detection of the decoherence due to spacetime fluctuations on
the Planck scale would provide experimental access to quantum
gravity effects analogous to accessing to atomic scales provided
by Brownian motion. A number of authors
\cite{PowerPercival2000,PercivalStrunz1997,Power1999} have
suggested investigating decoherence in the two-path atom
interferometer where the separation of the wavepackets is large
compared to the width of the wavepacket.   There have also been
studies of decoherence models using non-propagating conformal
fluctuations of spacetime by
Sanchez-Gomez~\cite{Sanchez-Gomez1993}, and using Newtonian
gravity by Kay~\cite{Kay1998}.

Recent work by Power and Percival~\cite{PowerPercival2000}
considered a model of the `conformal field' $A$ that conformally
deforms the Minkowski spacetime metric $\eta_{\alpha\beta} = $
diag$(-1,1,1,1)$ (using $\alpha,\beta=0,1,2,3$ as spacetime
coordinate indices)  into the curved spacetime metric
$g_{\alpha\beta}$ via
\begin{equation}\label{galbe0}
g_{\alpha\beta} = (1+A)^2 \eta_{\alpha\beta} .
\end{equation}
They assumed that the incoherent `conformal waves' of $A$ are
produced by their own quantum mechanical zero point fluctuations.
These waves then interact with wavepackets of massive particles
through the effective Newtonian potential using the weak field and
slow motion approximation. The nonlinear contribution from
conformal field causes a decoherence of the quantum wavepackets.
The random walk of the conformal field is modelled by a Gaussian
correlation function with $\tau_0$ as the correlation time. In
terms of the density matrix $\rho$ of the wavepacket at time $t=0$
and its change due to spacetime fluctuations $\delta\rho$ at time
$t=T$, the decoherence is represented by the loss of contrast of
the form~\cite{PowerPercival2000}:
\begin{equation}\label{decorhtot}
\frac{\delta\rho}{\rho(0)}
=
\sqrt{\frac{\pi}{2}}\frac{M^2 c^4 T
A_0^4 \tau_0}{\hbar^2}
\end{equation}
where $M$ is the mass of the atom and $A_0$ is the amplitude of
the fluctuating $A$, assumed to be much less than unity. The
conformal field is assumed to have frequency contributions from 0
up to a cut-off frequency $\omega_0 = {2\pi}/{\tau_0}$. Based on
atom interferometer experiments using caesium atoms with $M= 133$
amu separated for $T = 0.32$~s resulting in a loss of contrast of
$\delta\rho/\rho(0) = 3$\%~\cite{Peters1997}, Power and Percival
estimate the lower bound on $\lambda$ to be of  order 10, outside
the expected range.

Previous
models~\cite{PowerPercival2000,PercivalStrunz1997,Power1999}
suffer from a number of serious drawbacks and are too crude to
make predictions~\cite{Amelino-Camelia2005}. They suspend all but
the conformal degrees of freedom of gravity at the cost of
violating Einstein's equations. Consequently, most recent
investigations of gravitational decoherence have focused on
gravitational waves (GWs) of astrophysical and cosmological
origins~\cite{Reynaud2004, Lamine2006}. The smallness of these
waves have, however, rendered the chance of detecting their
decoherence effects very slim. Here we introduce for the first
time new components of the gravitational field describing shearing
actions on spacetime geometry, commonly referred to as the spin-2
GWs~\cite{Wang2005b, Wang2005c}. These components define the
conformal geometry of spacetime that yields the full geometry when
combined with the conformal field. Careful initial data analysis
reveals that the spin-2 GWs carry the true dynamical degrees of
freedom of GR~\cite{York1972}, just like the spin-1 EM waves
carrying the true dynamical degrees of freedom of Maxwell's theory
of electromagnetism. GWs are believed to be quantized into spin-2
gravitons and hence have zero point energy.

The essential requirement for the theoretical framework in which
the conformal filed interacts with GWs at zero point energy is a
conformally decomposed Hamiltonian formulation of GR. Such a
theoretical framework has been established in recent
papers~\cite{Wang2005b, Wang2005c}. It allows us to consider a
general spacetime metric of the form
\begin{equation}\label{galbe}
g_{\alpha\beta} = (1+A)^2 \gamma_{\alpha\beta} .
\end{equation}
in terms of the conformal field $A$ and the rescaled metric
$\gamma_{\alpha\beta}$. We shall work in a standard laboratory
frame where the direction of time is perpendicular to space.
Accordingly, we set $\gamma_{00} = -1$ and $\gamma_{0a} = 0$
(using $a,b=1,2,3$ as spatial coordinate indices.) The spatial
part of the metric $\gamma_{\alpha\beta}$ is denoted by
$\gamma_{ab}$ and is normalized using $\det(\gamma_{ab}) = 1$.
Hence, $\gamma_{ab}$ will be referred to as the `conformal metric'
as it specifies the conformal geometry of space. It's inverse is
denoted by $\gamma^{ab}$. The spacetime metric \eqref{galbe}
therefore accommodates both the conformal field, as does the
metric in \eqref{galbe0}, and in addition the spin-2 GWs encoded
in the deviation of the conformal metric $\gamma_{ab}$ from the
Euclidean metric $\delta_{ab}$.

Adopt, for a moment, units where $16\pi G = c = 1$. In
Refs~\cite{Wang2005b,Wang2005c}, the canonical theory of general
relativity has been constructed in terms of the conformal classes
of spatial metrics by extending the Arnowitt-Deser-Misner (ADM)
phase space consisting of the spatial metric ${g}_{a b}$ and its
momentum $p^{a b}$, ($a,b =1,2,3$). The canonical transformation
$({g}_{a b}, p^{a b}) \rightarrow (\gamma_{a b}, \pi^{a b};
\tau,\mu)$ is performed using a conformally transformed spatial
metric $\gamma_{a b}$, its momentum $\pi^{ab}$, the scale factor
$\mu =\sqrt{\det g_{ab}}$ and York's mean extrinsic curvature
variable $\tau$~\cite{York1972,Wang2005b}. The new canonical
framework has the Hamiltonian constraint $\H$, diffeomorphism
(momentum) constraint: $\D_a$ and conformal constraint
$\C$~\cite{Wang2005c}. In terms of these constraints, the
Hamiltonian for gravity is given by~\cite{Wang2005b,Wang2005c}:
\begin{equation}\label{Hfinal}
H = \int \left[ N \H + X^a \D_a + Z \,\C \right] \d^3 x
\end{equation}%
using the lapse function $N$, shift vector $X^a$ and Lagrange
multiplier $Z$ to effect the vanishing of the conformal constraint
$\C$. We choose coordinate and scaling gauge conditions to
facilitate the study of the coupling between the conformal and
gravitational wave parts of gravity. In addition, these gauge
fixings should allow our spacetime metric to be expressed in terms
of a deviation from the Minkowski metric for ease of comparison
with earlier approaches to the conformal field. We fix the
conformal gauge by normalizing the conformal 3-metric according to
$\det(\gamma_{ab}) = 1$. The momentum $\pi^{ab}$ is then required
to be traceless with respect to $\gamma_{ab}$. It is possible to
find parameterizations to satisfy the above properties of
$\gamma_{ab}$ and $\pi^{ab}$. However, their explicit construction
is not required for our present analysis.

The normalized conformal 3-metric $\gamma_{ab}$ can be extended
into a normalized conformal 4-metric $\gamma_{\alpha\beta}$ by
adopting the Gaussian coordinate conditions for the time-time and
time-space metric components $\gamma_{00}=-1$ and $\gamma_{0a}=0$.
Using the `conformal field' $A$, we can relate the physical and
conformal metrics simply by Eq. \eqref{galbe}. This fixes the
lapse function and shift vector to be $N = 1+A$ and $X^a = 0$
respectively, so that the direction of time is chosen as
perpendicular to space. We then perform the canonical
transformation $(\gamma_{a b}, \pi^{a b}; \tau, \mu) \rightarrow
(\gamma_{a b},\pi^{a b}; A, P)$ where $P$ is the momentum of $A$.
In terms of these variables, the gravitational Hamiltonian
\eqref{Hfinal} becomes
\begin{eqnarray}
H &=&
%\frac{c^4}{16\pi G}
\int \left[ \H^\text{(CF)} + \H^\text{(GW)} \right] \d^3x
\label{H}
\end{eqnarray}
where
\begin{eqnarray}
\H^\text{(CF)}
&=&
-
\left( \frac1{24}\, P^2 + 6\,\gamma^{ab}  A_{,a} A_{,b}
\right) \label{HHPhi}
\end{eqnarray}
is the Hamiltonian density for the conformal field and
\begin{eqnarray}
\H^\text{(GW)} &=&
%\frac{c^4}{16\pi G}
%\left(
(1+A)^{-2}\pi{}_{ab}\pi^{ab} - (1+A)^2  R_\gamma
%\right)
\label{HHgamma}
\end{eqnarray}
is the Hamiltonian density for the GWs, where $R_\gamma$ is the Ricci scalar curvature of
$\gamma_{ab}$.
Although we have adopted a canonical description, the covariance of our approach can be reaffirmed
by means of a Legendre transformation on the Hamiltonian
\eqref{H}, yielding the action for gravity of the covariant form:
\begin{eqnarray*}
S &=&
\int
\left[
6\,\gamma^{\alpha\beta} A_{,\alpha}  A_{,\beta}
+
(1+A)^2 \R_\gamma \right] \d^4x
\end{eqnarray*}
where $\R_\gamma$ is the Ricci scalar curvature of the 4-metric
$\gamma_{\alpha\beta}$. In arriving at this action, we have used
the momentum-velocity relations
$\dot{\gamma}_{ab} = 2 \,(1+A)^{-2}\pi_{ab}$ and
$\dot{A} = -P/12$, where the overdot denotes a time derivative.
By using the latter relation, we can eliminate the momenta in
$\H^\text{(CF)}$ and write it, after restoring the full units, as
\begin{eqnarray}
\label{HCF}
\H^\text{(CF)}
&=&
-\frac{3 c^4}{8\pi G}
\left(
c^{-2}\dot{A}^2
+
\gamma^{ab} A_{,a} A_{,b}
\right) .
\end{eqnarray}
This Hamiltonian density has a remarkable feature of being similar
to that of a massless scalar field but with a `wrong sign', i.e.
negative energy density, which has important physical consequences
to be explored below.

It is worth stressing that only physical degrees of freedom have
zero point energies. In principle, 3 conditions can be used to
eliminate the coordinate redundancy in the conformal metric
$\gamma_{ab}$ containing 5 components.
Assuming small deviation from the Minkowski spacetime, the vacuum
energy due to $\H^\text{(GW)}$ in the semiclassical domain can be
estimated by considering the quantization of weak GWs in a box of
unit volume. In this case, the number of modes per frequency is
given by
\begin{equation}\label{numodes}
n(\omega) = \frac{4\pi\omega^2}{(2\pi c)^3} .
\end{equation}
The energy
density $\H^\text{(GW)}_0$ due to zero point energy gravitons with
2 polarizations and frequencies up to the cut-off value $\omega_0 = {2\pi}/{\tau_0}$
is
\begin{equation}\label{E0}
\H^\text{(GW)}_0 \approx 2 \int_0^{\omega_0} \frac12\hbar\omega\,
n(\omega)\d\omega
= \frac{2\pi^2
c^2}{\lambda^4}\,\frac{M_\text{Planck}}{L_\text{Planck}^3}
\end{equation}
since each frequency mode has energy $\frac12\hbar\omega$. For
$\lambda > 10^2$, the above
$\H^\text{(GW)}_0$ value amounts to a vacuum mass density of
$10^{68}$--$10^{84}$ metric tons per litre! However, we must also take into
account the contribution from the conformal field, which can
be determined using the Hamiltonian constraint to be:
\begin{equation}\label{Acond}
\H^\text{(CF)} = -\H^\text{(GW)}_0 .
\end{equation}
The relation \eqref{Acond} admits a very simple picture: Every pair of
gravitons having the same zero point energy $\frac12\hbar\omega$
but two different helicities is accompanied by a `quantum' of
the conformal field with a compensating energy of $-\hbar\omega$.
However, a further implication is that the resulting conformal
field may be observable through an atom wave decoherence
experiment. Since all components of the metric deviating from the
Minkowski values are assumed to be small, the Newtonian
approximation for the slow motion of a matter wave packet applies.
However, in this limit, only the conformal field contributes to
the effective Newtonian gravitational potential $-(g_{00}+1)/2$
but not the GW fields. The amplitude of the fluctuating conformal
field $A $ can be derived from the energy density condition
\eqref{Acond} together with Eqs \eqref{HCF} and \eqref{E0} in a
mode-by-mode fashion as follows:
\begin{equation*}%\label{}
\frac{3c^4}{8\pi G}
\ave{
c^{-2}\dot{A}^2
+
\abs{\nabla A}^2} \d\omega
=
\hbar\omega n(\omega)\d\omega
\end{equation*}
where $\ave{\cdot}$ denotes averaging over space and the vector
modulus $\abs{\cdot}$ is approximated using the Euclidean norm.
Let $A(\omega)$ be the amplitude of $A$ at frequency $\omega$.
Using relation \eqref{numodes}, the above relation yields
$\ave{A(\omega)^2}={2}\,T_\text{Planck}^2\omega/{3\pi}$ which
shows a clear spectral distribution of the conformal field towards
higher frequencies. From this, the overall amplitude squared for
$A$ due to quantum vacuum fluctuations is
\begin{equation}\label{aveAsqtot}
A_0^2
=
\int_0^{\omega_0}\ave{A(\omega)^2}\d\omega
=
\frac{4\pi}{3\lambda^2} .
\end{equation}
Within the Newtonian approximation for the atom
wave motion, we finally obtain an expression for the cut-off parameter $\lambda$
by substituting Eq.~\eqref{aveAsqtot} into Eq.
\eqref{decorhtot} to be
\begin{equation}\label{lambdaBMW}
\lambda
=
\left(
\frac{8\sqrt{2\pi^{5}}}{9}
\frac{M^2 c^4 T_\text{Planck} T}{\hbar^2(\delta\rho/\rho(0))}
\right)^{1/3} .
\end{equation}
For the mentioned atom interferometer experimental data using
caesium atoms with $M = 133$ amu separated for $0.32$~s resulting
in a loss of contrast of about 3\%~\cite{Peters1997}, formula
\eqref{lambdaBMW} yields a lower bound on $\lambda$ to be:
\begin{equation}\label{lamb7600}
\lambda \ge 7600 .
\end{equation}
The inequality counts for any other causes of decoherence. This
lower bound is well within the expected range $\lambda > 10^2$ for
low energy quantum gravity.

For almost a century it has been widely perceived that the lack of
experimental evidence for quantum gravity has presented, and will
continue to present, a major barrier to its breakthrough. However,
armed with the sensitivity of modern matter wave interferometers
at the quantum level, the possibility of using a `macroscopic'
instrument to investigate Plank scale physics is now a real
possibility. Following recently formulated conformal decomposition
in full canonical gravity, we have developed a new approach to
gravitational decoherence due to ground state gravitons and have
demonstrated that the resulting conformal field can lead to
observable effects by causing quantum matter waves to loss
coherence.

Further improved measurement may decrease the upper bound of
decoherence resulting in an increased $\lambda$. A space mission
similar to the proposed HYPER atom wave interferometer can provide
such improvements~\cite{hyper}. Advanced matter interferometers
using molecules under active
development~\cite{Hornberger2003,Hackermueller2003,Hackermueller2004}
are designed to enhance decoherence effects using particles of
larger mass. However, the additional decoherence effects due to
the structure of a molecule must be controlled. Recent experiments
using C$_{70}$ fullerene molecules with $M = 70\times12$ amu,
separated for $0.004$~s result in a loss of contrast of
4\%~\cite{Hackermueller2004}. According to formula
\eqref{lambdaBMW}, this yields $\lambda\ge 5500$, close to the
lower bound given in Eq.~\eqref{lamb7600}. For every 3 orders of
magnitude improvement in the loss of contrast, there is only 1
order of magnitude change in the lower bound of $\lambda$.
Therefore, its possible rise with any decrease of decoherence
measurement is rather slow. This, combined with the fact that the
lower bound of $\lambda$ in the range 5500--7600 calculated using
current experimental data is already within the expected range
$\lambda > 10^2$, is a very good sign. It strongly suggests that
the measured decoherence effects are converging towards the
fundamental decoherence due to quantum gravity. The experimental
determination  of $\lambda$ will be compelling evidence for the
quantum behaviour of spacetime and set a stringent benchmark in
our search for the ultimate theory of quantum gravity.

\section*{Acknowledgments}
We would like to thank M Sandford, G Amelino-Camelia, K
Hornberger, B Kent, I C Percival and J W York for fruitful
discussions.  The work was supported by the CCLRC Centre for
Fundamental Physics.

%\newpage

\end{document}